# Federal Learning Framework for Quality Evaluation of Blastomere Cleavage

Jung-Hua Wang
AI Research Center
Department of Electrical Engineering
National Taiwan Ocean University
Keelung 20224, Taiwan
jhwang@email.ntou.edu.tw

Huai-Wen Chang
Department of Electrical Engineering
National Taiwan Ocean University
Keelung 20224, Taiwan
huaiiwenn@gmail.com

Rong-Yu Wu
Department of Electrical Engineering
National Taiwan Ocean University
Keelung 20224, Taiwan
abner90635@gmail.com

Ting-Yuan Wang
Industrial Technology Research Institute
Hsinchu County 310401, Taiwan
vpwang@gmail.com

Ming-Jer Chen
Division of Infertility
Lee Women's Hospital
Taichung 40705, Taiwan
mingjerchen@gmail.com

Yu-Chiao Yi
Division of Reproductive
Endocrinology and Infertility
Veterans General Hospital (VGHTC)
Taichung 40705, Taiwan
yuchiaoyi@gmail.com

*Abstract* —This study addresses the issue of leveraging federated learning to improve data privacy and performance in IVF embryo selection. The EM (Expectation-Maximization) algorithm is incorporated into deep learning models to form a federated learning framework for quality evaluation of blastomere cleavage using two-dimensional images. The framework comprises a server site and several client sites characterized in that each is locally trained with an EM algorithm. Upon the completion of the local EM training, a separate 5-mode mixture distribution is generated for each client, the clients' distribution statics are then uploaded to the server site and aggregated therein to produce a global (sharing) 5-mode distribution. During the inference phase, each client uses image classifiers and an instance segmentor, assisted by the global 5-mode distribution acting as a calibrator to (1) identify the absolute cleavage timing of blastomere, i.e., tPNa, tPNf, t2, t3, t4, t5, t6, t7, and t8, (2) track the cleavage process of blastomeres to detect the irregular cleavage patterns, and (3) assess the symmetry degree of blastomeres. Experimental results show that the proposed method outperforms commercial Time-Lapse Incubators in reducing the average error of timing prediction by twofold. The proposed framework facilitates the adaptability and scalability of classifiers and segmentor to data variability associated with patients in different locations or countries.

## INTRODUCTION AND RELATED WORK

In vitro fertilization (IVF) and intracytoplasmic sperm injection (ICSI) are two mainstream means in assisted reproductive technologies (ART) for treating infertility. The selection and transfer of embryos are critical steps that determine the success of ART and directly impact clinical outcomes. Therefore, assessment of embryo quality plays a crucial role in infertility treatment. Recently, federated learning (also referred to as collaborative learning) has drawn great interest from the research community of reproductive medicine due to its decentralized nature in training machine learning models and requiring no exchange of data from client devices to global servers. Client data on edge computing devices is used to train models locally, improving data privacy and data sovereignty, which is particularly important in the field of reproductive medicine.

Embryo quality assessment has been a hot research topic in the reproductive medicine field for many years. A study by Alikani et. al. (1999) found that large fragments are harmful to developing embryos, while localized or small, scattered fragments do not significantly affect embryos. Embryos with lower degrees of fragmentation have higher implantation and pregnancy rates. Hardarson et. al. (2001) found that by comparing blastomere size after cleavage, uneven cleavage (a form of asymmetry) resulted in higher levels of aneuploidy or multinucleation in the embryo. This could reduce the implantation rate and harms the pregnancy outcome of IVF. Bączkowski et. al. (2004) reviewed various traditional methods for evaluating embryo quality and found that both predicting embryo developmental potential and the likelihood of pregnancy are closely related to morphological and dynamic characteristics. They also proposed a new embryo grading standard based on the number of blastomeres, blastomere symmetry, and the degree of blastomere fragmentation.

Good quality embryos typically have 4 to 6 cells on the second day and 8 to 12 cells on the third day. In addition, the higher the consistency of blastomere size and the lower the degree of fragmentation, the better the embryo quality. Another study by Herrero et. al., (2013) showed that the timing of cleavage in cleavage-stage embryos is closely related to their developmental potential and quality. Early cleavage can lead to abnormal division of genetic material, causing aneuploidy, while late cleavage may indicate DNA damage or chromosomal abnormalities. Embryos that cleave at appropriate time points represent intact cytoplasmic components and have a higher chance of becoming blastocysts. Therefore, cleavage timing is an important indicator for embryo quality assessment, which significantly affects the success rate of infertility treatment and patient pregnancy outcomes. Another key influential topic in blastomere development is irregular cleavage (IRC) patterns, which were thoroughly studied by Yang, et al. (2015). They found that some specific IRC patterns can reduce the potential of embryos to develop into high-quality blastocysts and predict blastocyst formation and quality based on cleavage patterns.

All of the above studies relied on static microscopy images. Before the widespread use of time-lapse incubators (TLI), imaging data in infertility research mainly consisted of optical microscope images, capturing only specific time points of embryo morphology. The lack of sequential imaging data results in incomplete monitoring of embryonic development, and critical changes or abnormalities can easily be missed due to heavy reliance on subjective manual operations and analysis. While rapidly advancing machine learning can alleviate the lack of automation, most commercial AI-driven TLIs are costly and mostly closed systems trained with undisclosed big datasets, making their predictions not always interpretable and reliable (to be seen later). Consequently, how to train an AI model with limited imaging data collected from small medical institutions or clinics without referring to patients' pathological data to provide reliable morphokinetic parameters that characterize the cleavage of blastomeres is a topic worthy of study.

Instead of using static image, Meseguer, et al. (2011) performed a tree-based statistical analysis on a sequence of images containing morphological and dynamic parameters of embryos. It started with morphological screening by excluding non-viable embryos, followed by those with inconsistent blastomere size at the 2-cell stage, direct cleavage, and multinucleation at the 4-cell stage. Finally, by defining three timing parameters, namely $t5$ (48.8-56.6 hours), $s2$ (≤0.76 hours), and $cc2$ (≤11.9 hours) as the primary, secondary, and third splitting nodes, embryos were classified into ten categories for predicting implantation rates. Other related research works by Motato et. al. (2016) and Cetinkaya, et. al. (2015) also used these blastomere characteristics to predict embryo developmental potential, including the likelihood of developing into high-quality blastocysts, implantation rate (Sela, et. al. 2012), and pregnancy rate (Kirkegaard, et. al. 2013). In general, absolute cleavage timing, blastomere symmetry, fragmentation, and cell-division status are key morphokinetic parameters affecting embryo quality and can be used for embryo quality assessment.

Table 1 Schematic diagram of cell stages

| Cell Stage | 1-cell stage | 2-cell stage | 4-cell stage | |
|---|---|---|---|---|
| Number of Blastomere | 1 blastomere | 2 blastomeres | 3 blastomeres | 4 blastomeres |
| Time-lapse Image | | | | |
| Cell Stage | 8-cell stage | | 8-cell stage | |
| Number of Blastomere | 5 blastomere | 6 blastomeres | 7 blastomeres | 8 blastomeres |
| Time-lapse Image | | | | |

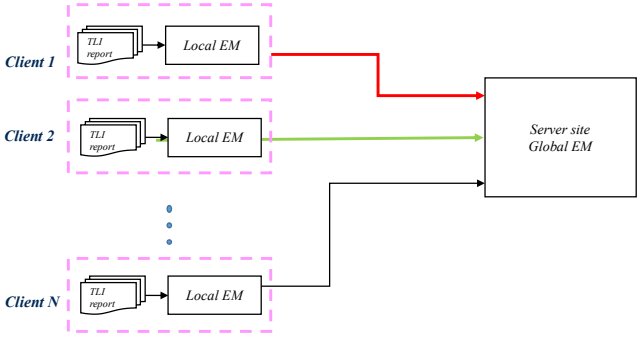

Fig. 1 Architecture of the proposed framework.

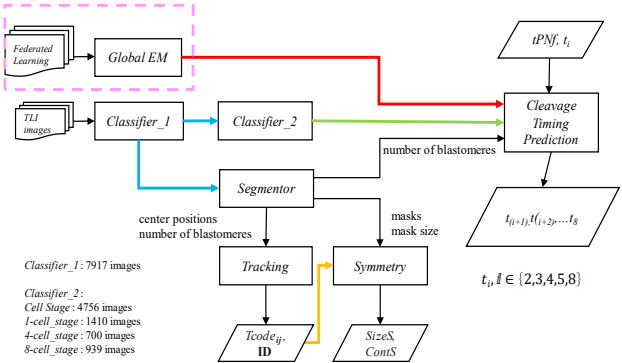

Fig. 2 The inference flowchart at the Client site.

Based on deep learning, a study by Liao, et al. (2021) trained a DenseNet201 (Huang et. al. 2017) to classify blastomeres based on cell number into five categories: 1-cell stage, 2-cell stage, 3-cell stage, 4-cell stage, and ≥5-cell stage. However, their results can only be used to predict blastocyst formation and quality because only PN (pronucleus) initiation time was given and cleavage times for other cell stages were not quantified. Another study (Sharma, 2023) used object detection methods to classify stages from 1-cell to 9+-cell, morula, and blastocyst, but the proposed method could only obtain the start time of the 5-cell stage with a delay of 2-3 hours. Although the aforementioned embryo characteristics all affect embryo quality, including images with a high degree of fragmentation in the feature quantification system severely impacts the accuracy of extracting other features of the embryo; embryos with a low degree of fragmentation have little effect on implantation potential.

Despite that Hall, et. al. (2024) proposed the use of federated learning to develop an AI model for predicting usable blastocyst from pre-ICSI oocyte images, so far there is no literature report of using federated learning in predicting ACT and mophorkinetic parameters quantification. In light of these observations, our goal is to develop a low-cost federated learning framework that involves deep networks and the EM algorithm trained on 2-D TLI images to build a low-cost, decentralized, and highly interpretable AI model for assessing a cleavage stage embryo. Such a model is capable of quantifying blastomere cleavage timing, blastomere symmetry, and cell-division status through the incorporation of image classification, image segmentation, and Expectation-Maximization algorithm (EM) developed by Moon, (1996), we aim to provide ancillary recommendations for reaching clinical decisions in embryo selection and infertility research. Table 1 shows the conventional cell stages (i.e., 1-cell, 2-cell-, 4-cell, and 8-cell) defined in terms of the number of blastomeres during the cleavage stage. Notably, the 8-cell stage covers the time interval of the presence of 5, 6, 7, 8 blastomeres.

## METHODOLOGY

This paper presents a federated learning framework aiming to provide more complete, compared to commercial TLIs, characterizations of morphokinetic parameters for embroys at the cleavage stage. Fig. 1 shows the proposed framework for achieving this goal, where a sequence of embryo images is fed into each client and subject to the training by the EM algorithm to obtain a set of 5-mode distribution that fits the statistics of input images. The five modes refer to the Gaussian distributions of $t_2$, $t_3$, $t_4$, $t_5$, and $t_8$. After the local training, the resultant statics parameters such as mean and deviation are sent to the server site and therein are subjected to a global EM training to obtain the final set of the 5-mode distributions. These shared parameters will be used by each client site, which is presumably installed at a different hospital or clinic, to conduct the inference task of determining the absolute cleavage timing for each image frame.

Fig.2 shows how the client site shares the global EM-trained parameters, in conjunction with image classification and instance segmentation, to complete the inference task for the eight absolute timings. *Classifier_1* for selecting images at the cleavage stage, and *Classifier_2* takes the selected images from *Classifier_1* to perform classification according to the number of blastomeres. The output of *Classifier_1* is also sent to a *Segmentor* for instance segmentation, the resulting masks can be used in several aspects. The number of masks can not only be used to check if it is in line with the prediction of *Classifier_2*, the output masks can also be used to compute the blastomere area, and the center positions thereof. The position information is sent to the Tracking block, which tracks the cleavage process of blastomeres. The tracking result and mask information are further sent to the Symmetry block, which executes the symmetry evaluation (in terms of similarity in mask area and contour) between blastomeres that are subjected to the same number of cleavage cycles. On the other hand, a set of ideal report data that encompasses all absolute cleavage timing (ACT) points provided by the TLI instrument is used to train the EM algorithm to obtain the mean and deviation and hence the normal distributions of each ACT timepoint. The process of Cleavage Timing Prediction (CTP) first compares the timepoint class prediction of *Classifier_2* and that of *Segmentor* to obtain tPNf and $t_i$, $i \in$

{2, 3, 4, 5, 8}, if they are inconsistent, then the aforesaid normal distributions are used to calibrate the aforesaid timepoints, thereby determining $t_1$, $t_1= t_{(i+1)}$, $t_{(i+2)}$, ⋯ $t_8$. Details of the calibration are elaborated in the following discussions.

The dataset used in this study was collected from the report data (including images each taken every 15 mins) of EmbryoScope™ at Taichung Veterans General Hospital (VGHTC) between 2021 and 2023. We conducted a data cleaning process to ensure high-quality data for our study and make it suitable for model training and analysis. The retrospective images were annotated by the physician coauthors. The first step in data cleaning is to examine Gardner-graded (Gardner, et. al. 1999) embryos while excluding TLI images with high fragmentation degrees, dark and blurry, and removing report data with incomplete or outlier cleavage timing. To verify the feasibility of our idea, it is sufficient to split the cleaned images into three subsets, not necessarily equalized, to simulate multiple clients having a different dataset size. In Fig.2, each client is assigned with a split dataset for training deep network models of ResNet50 (He, et.al. 2016) and YOLOv5 (Jocher, et al. 2020). In total, the ResNet50 has 4894 training images, 1400 validation images, and 700 test images. The YOLOv5 has 333 training images, 112 validation images, and 50 test images. A set of ideal report data collected from 910 embryos that encompass all ACT points for tPNf, $t_2$, $t_3$, $t_4$, $t_5$, and $t_8$ is split and used for training the local EM algorithm. Each image is standardized to a resolution of 500x500 pixels. After the local EM algorithm, each client site will have a separate normal distribution for $t_2$, $t_3$, $t_4$, $t_5$, and $t_8$. Key statistical parameters of these distributions are then sent to the server for producing an aggregated 5-mode distribution for sharing use during the subsequent inference task of all client sites (see Fig.2). The aggregation can adopt the strategy of Max pooling or weighted average pooling.

According to *Article 12*, *paragraph 2* of **Human Subjects Research Act (Taiwan)** that states "⋯ But the research protocol within the scope of exemption categories for consent requirements, as announced by the competent authority, shall not apply", the dataset collection in this work falls within the scope of exemption and fully complies with the second and third items of the act.

**Applying the EM Algorithm for Time Referencing**

For each client site, the (see pink dashed box in Fig. 1). First, the ACT timepoints of ideal report data, i.e., $t_2$, $t_3$, $t_4$, $t_5$, and $t_8$, are all subtracted by tPNf. These adjusted times points w.r.t tPNf serve as the training data for the EM algorithm. Since these data points are unlabeled, a plausible assumption is that they originate from five Gaussian models. Note that all of them are single modal, except that the model for t3 is a 2-mode mixture. The goal here is to estimate these Gaussian distributions automatically. The EM algorithm starts by initializing (e.g. averaging over all samples) parameters of mean and variance for five models. During the E-step, we first calculate the likelihood of each cleavage time point belonging to a particular Gaussian model using the initialized probability density functions. After the local training, the parameters of each five modes are sent to the server site and aggregated there by another EM training algorithm to produce a final set of parameters. In this way, these parameters can fit all the datasets from hospitals that participate the federated learning.

Then, assume the prior probability of cleavage time point associated with each Gaussian model is equal, we apply Bayesian formula to calculate the posterior probability of cleavage time point for each Gaussian model. During the M-step, these posterior probabilities are used to the parameters of each Gaussian model, recalculating the update mean and variance to ensure a better fit with the observed data. The algorithm iterates through the E-step and M-step, continuously updating the models until convergence. This iterative process ultimately yields five Gaussian models fitting the distribution of cleavage time points, ensuring robustness and precision of the final results and providing a reliable representation of timepoint distributions.

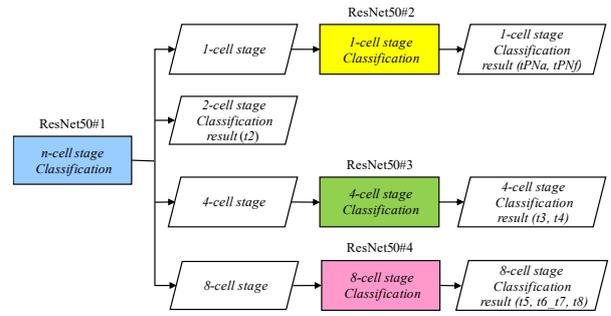

Fig. 3 Flowchart of *Classifier_2* that uses prediction outputs of *Classifier_1* to perform classification.

One should note that the EM method can yield the normal distribution models for $t_2$, $t_3$, $t_4$, $t_5$, and $t_8$ specific to patients, along with their mean and variance, using a small training dataset. Additionally, it allows for adjusting these normal distribution models while keeping their time intervals (e.g. the interval between $t_5$ and $t_8$) intact, enabling the models to adapt to noises or fluctuations in the embryo image characteristics. This property underscores the adaptability of our method. Since the initial cleavage time distribution for $t_3$ is modeled as a Gaussian mixture model based on VGHTC patients' data, we can simply set the average initial cleavage time for $t_3$ as the mean of $t_4$ minus one standard deviation. The standard deviation for $t_3$ is set to half of the standard deviation of $t_4$. In CTP, the values of tPNf and t2 are returned to the EM algorithm (see the upper red arrow in Fig.1), whereby the average value of the ACT normal distributions gets adjusted. This adjustment is useful in calibrating the bias error because the EM algorithm was pre-trained with the TLI's report data, it literally fits the distributions of $t_2$, $t_3$, $t_4$, $t_5$, and $t_8$ to the true distribution of embryos collected in VGHTC, thereby enhancing the accuracy of blastomere CTP.

*Classifier_1*

In reproductive medicine, main stages of embryo development are: Cleavage-stage, Morula-stage, and Blastocyst-stage. This study focuses solely on the characteristics of Cleavage-stage embryos. As shown in Fig. 1, after the classification by *Classifier_1*, from the input sequence of TLI images, a subset of images that contain one or more blastomeres will be selected and sent to *Classifier_2* for subsequent analysis.

*Classifier_2*

In Fig. 3, *Classifier_2* is responsible for giving one of the eight categories as its output: tPNa, tPNf, $t_2$, $t_3$, $t_4$, $t_5$, $t_6\_t_7$, and $t_8$. To achieve this goal, we employ four image classifiers (e.g. RestNet50) trained separately for performing a single class task. The first ResNet50 classifies the input image into four classes: 1-cell stage, 2-cell stage, 4-cell stage, and 8-cell stage. Afterwards, the second ResNet50 takes 1-cell stage input image and classifies it as either tPNa or tPNf. The third ResNet50 takes the 4-cell stage image and classifies it as either t3 or t4. Finally, the fourth ResNet50 classifies the 8-cell stage image into $t_5$, ($t_6\_t_7$), and $t_8$.

**Instance Segmentation**

*Segmentor* will take the output from *Classifier_1* to perform instance segmentation, thus obtaining the number of blastomeres as well as blastomere-related parameters such as center positions, and areas. Before proceeding, we first note that object detection can only provide the number of blastomeres because it does not have the mask information as in instance segmentation. Instance segmentation, on the other hand, not only provides the number of blastomeres but also yields additional segmentations that allow us to perform the tasks of tracking blastomere cleavage and blastomere symmetry calculation, which are essential to assess the quality of the embryo. In short, instance segmentation can offer more comprehensive data at the expense of greater complexity in the pixel-by-pixel annotation.

TABLE 2 Coding scheme for tracking

| $Tcode_{ij}$ | ID | $Tcode_{ij}$ | ID |
|---|---|---|---|
| (0,) | zygote | (0, 0, 0, 0) | A00 |
| (0, 0) | A | (0, 0, 0, 1) | A01 |
| (0, 1) | B | (0, 0, 1, 0) | A10 |
| (0, 0, 0) | A0 | (0, 0, 1, 1) | A11 |
| (0, 0, 1) | A1 | (0, 1, 0, 0) | B00 |
| (0, 1, 0) | B0 | (0, 1, 0, 1) | B01 |
| (0, 1, 1) | B1 | (0, 1, 1, 0) | B10 |
|  |  | (0, 1, 1, 1) | B11 |

**Pseudo code for tracking blastomere cleavage**

Input: $\{CL_{ij}, ID_{ij}\}$, $i$=1, 2, ... $k$, $j$=1, 2, ... $l$
Output: $\{Tcode_{ij}\}$
1  for $i = 1:k$
2      for $j = 1:l$
3          if $len(CL_{ij}) \geq 1$:
4              if $len(CL_{(i-1)j}) == len(CL_{ij})$:
5                  $Tcode_{ij}$
                   $= Function1(CL_{(i-1)j}, CL_{ij}, ID_{(i-1)j})$
6              elif $len(CL_{(i-1)j}) != len(CL_{ij})$:
7                  $Tcode_{ij}$
                   $= Function2(CL_{(i-1)j}, CL_{ij}, ID_{(i-1)j})$

To address the overlapping nature of blastomeres during the cleavage stage, we adopted a quick and effective strategy of raising the threshold of IOU (Intersection over Union) for Non-Maximum Suppression (NMS). Heuristically, 0.65 works fine for most cases. The beauty of doing so is that the accuracy of segmenting overlapping blastomeres can be increased without additional training data.

**Tracking the Cleavage Process**

In practice, embryo cleavage is an important indicator of embryo quality. Since observing embryo cleavage is very time-consuming, in this work we developed a Tracking scheme (see Pseudo code below) to track the cleavage process and highlight embryos with irregular cleavage. To this end, we design a coding scheme named $Tcode_{ij}$ (Table 2) for tracking an arbitrary blastomere, the result of which can be used to perform symmetry evaluation. The index $i$ denotes the number of TLI images, $j$ is the number of blastomeres in each TLI image, and CL represents the center position of every blastomere. $ID_{ij}$ is the sequence number of every blastomere in each TLI image. We have to check whether the number of blastomeres in the previous image $CL_{(i-1)j}$ is equal to that in the current image $CL_{ij}$. Based on the checking result, the tracking process can only have two distinct cases: (1) $CL_{(i-1)j}$ and $CL_{ij}$ are the same, this indicates that the blastomeres have not undergone division, and the number of blastomeres remains constant. Here, we input $CL_{(i-1)j}$, $CL_{ij}$, and $Tcode_{(i-1)j}$ into Function1 for matching the blastomeres by finding those with the closest center position between consecutive images, and for updating the $Tcode_{ij}$ to ensure that the same blastomere in consecutive images retains a consistent $Tcode_{ij}$, thereby indicating continuity. (2) $CL_{(i-1)j}$ and $CL_{ij}$ are different, this signifies that blastomeres have undergone division. In that case, $CL_{(i-1)j}$, $CL_{ij}$, and $Tcode_{(i-1)j}$ are fed into Function2 for identifying the divided blastomeres and assigning $Tcode_{ij}$ according to the division status of the blastomeres.

Function2 is explained as follows: $Tcode_{ij}$ is assigned according to Table 2. Usually, the cleavage process starts with one blastomere, $Tcode_{ij}$ is (0) and ID = zygote. After the first cleavage, $Tcode_{ij}$ is updated into (0,0) and (0,1) for the two separate blastomeres. The IDs of (0,0) and (0,1) are A and B, respectively (marked red in Table 2). Following this coding rule, the updated code is derived by appending 0 or 1 to the pre-cleavage code. This allows for tracking each cleavage event and facilitates embryology focus on analyzing irregular cleavage embryo information.

$$SizeS = \frac{1}{1+\sigma} \times 100 \qquad (1)$$

**Quantizing the degree of Symmetry**

Fig. 2 shows that the Symmetry scheme uses the blastomere mask, and blastomere mask size as input data. The blastomeres to be compared are selected according to the $Tcode_{ij}$ provided by the Tracking. The length of a $Tcode_{ij}$ indicates the number of cleavage cycles a blastomere has undergone, while the binary code indicates from which original blastomere the current blastomere has divided. By comparing blastomeres with the same length of $Tcode_{ij}$, we can assess the symmetry at the 1, 2, 4, and 8 cell stages. For instance, (0,0,0,0), (0,0,0,1), (0,0,1,0), (0,0,1,1) have the same length (n = 4), and (0,0,0,0) and (0,0,0,1) have the same (n−1) binary digits (marked blue in Table 2). By comparing blastomeres with the same code length and the same binary digits, we can determine the symmetry of the original blastomere cleavage. With the coding, Symmetry is evaluated by comparing the similarity of area size and contour of the blastomeres. We use two metrics: $SizeS$ and $ContS$. To calculate $SizeS$, the standard deviation $\sigma$ of the blastomere mask size of the selected blastomeres is plugged into (1). For $ContS$, we invoke the well-known function of cv2.matchShapes in OpenCV. For convenience, we have normalized the returned value [0,1] of $ContS$ to [0,100]. The higher values of these two metrics suitably represent the greater degree of symmetry among the blastomeres.

**Cleavage Timing Prediction (CTP)**

This section discusses how to determine tPNf, $t2, t3, t4, t5, t6, t7$, and $t8$. First, we identify the cleavage timings for $tPNf$ and $t_i$ using outputs from Classifier_2 and the Segmentor, where $t_i$ represents the next ACT class recognized after $tPNf$. Note that the initial $t_i$ is not necessarily $t_2$, it could be $t_4$ or others as we might encounter direct cleavage. The value of $t_i$ is calibrated using Eq.(1) through Eq.(3). First, the value of $t_i$ is subtracted by $tPNf$, which is confirmed by Classifier_2 and segmentor, to obtain $R_{ti}$. A difference value $D_{ti}$ used for calibration is calculated by subtracting $R_{ti}$ from the parameters $\mu_{EMti}$ provided by the federated learning (from the global EM in Fig.1)

$$R_{ti} = ti - tPNf, R_{ti} = \mu_{ti} \qquad (1)$$
$$D_{ti} = R_{ti} - \mu_{EMti} \qquad (2)$$

Finally, Eq.(3) calibrates the adapted mean values of ACT distributions. Clearly, the calibration in effect eliminated the noise or fluctuations existent in the input frame, so as to allow an individual embryo to statistically adapt to its population, thereby boosting the prediction accuracy through federated learning.

$$\mu_{t2} = \mu_{EMt2} + D_{ti} \qquad (3)$$
$$\mu_{t3} = \mu_{EMt3} + D_{ti}$$
$$\mu_{t4} = \mu_{EMt4} + D_{ti}$$
$$\mu_{t5} = \mu_{EMt5} + D_{ti}$$
$$\mu_{t8} = \mu_{EMt8} + D_{ti}$$

In this way, the relative interval between timing points, especially $cc2$=t3- t2, $cc3$= t5- t4, $s2$= t4- t3, and $s3$= t8- t5 which are widely adopted norm for evaluating quality of embryo, are keep unchanged while fitting individual embryos. Because in this work each input frame in Fig.2 is designed to be jointly detected by Classifier_2 and Segmentor for a more stable prediction, it is important to check the consistency of the predictions between them. If they are inconsistent to each other, the EM pretrained ACT and the distributions of all timepoints are used as an arbitrator to decide whether to trust the prediction output from Classifier_2 or from Segmentor. Each image frame of the input sequence will be classified as one of the seven classes, $t_2, t_3, t_4, t_5, t_6, t_7$, and $t_8$. As the last step, Optical Character Recognition (OCR) is applied to the first frame within each class to identify the timing value (in hrs) stamped by TLI, thereby determining the absolute cleavage timing of tPNf, $t_2, t_3, t_4, t_5, t_6, t_7$, and $t_8$.

## Experimental Results

**Experiment Setup**
Hardware: CPU i9-9900 with on NVIDIA GPU/RTX-2080 and RAM/32G. Software: Windows10, Python 3.8, Cuda 11.1, cuDNN 8.0.5.

Fig. 4(a) shows the raw TLI images of overlapping blastomeres, Fig. 4(b) illustrates the segmentation result of overlapping blastomeres when the IOU threshold was set to 0.60 for IOU for NMS, which missed one blastomere, and Fig. 4(c) shows the segmentation result of overlapping blastomeres when the IOU threshold was set to 0.65 for NMS, which obtain the correct number of blastomeres in TLI images. Fig. 5 illustrates the process of normal cleavage and symmetry analysis results. Through the ID (see TABEL II), which shows in the center blastomere, the cleavage process can be known, and identify which blastomere each one originated from. At the same time, the blastomeres symmetry comparison result is shown in the top left corner of the TLI images. The symbol *SizeS* means the blastomeres area similarity and the *ContS* means the blastomeres contour similarity, higher scores indicate greater similarity and, thus, more symmetry among the blastomeres. Fig. 5(a) shows the blastomere that has not yet undergone cleavage because there is only one blastomere in the images, so we don't need to estimate the symmetry for the blastomere. Fig. 5(b) shows two blastomeres (**ID**: A, B) and (*SizeS_2, ContS_2*)= (82%, 73%), where *Tcode* length=2, which means both the two blastomeres have undergone one cleavage cycle. Fig.5(c) shows three blastomeres with (**ID**: A, B0, B1) and (*SizeS_3, ContS_3*)=(93%, 97%). Fig.5(d) shows four blastomeres with (**ID**: A0, A1, B0, B1) and (*SizeS_3, ContS_3*)= (76%, 89%). Fig.5(e) shows five

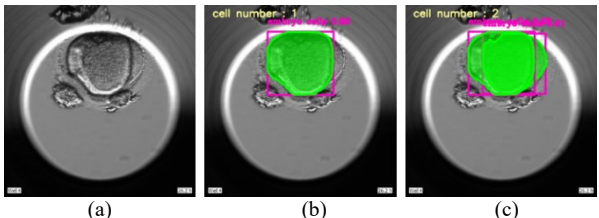
(a)      (b)      (c)
Fig. 4 I(a) raw TLI image (b) IOU threshold=0.60 resulted in 1 blastomere; (c) IOU threshold=0.65 resulted in two blastomeres.

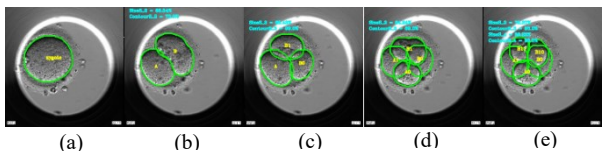
(a)    (b)    (c)    (d)    (e)
Fig.5 (a) 1 blastomere; (b) 2 blastomeres with (*SizeS_2, ContS_2*)= (82%, 73%); (c) 3 blastomeres with (*SizeS_3 ContS_3*)=(93%, 97%); (d) 4 blastomeres with (*SizeS_3, ContS_3*)=(76%, 89%); (e) 5 blastomeres with (*SizeS_3, ContS_3*)= (87%, 93%) and (*SizeS_4, ContS_4*)= (97%, 93%).

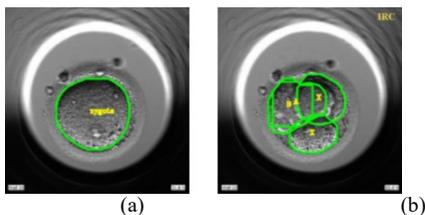
(a)        (b)
Fig. 6 Irregular cleavage. (a) zygote (b) The zygote cleaves directly into four blastomeres, labeled with "IRC" in the top right corner, showing irregular cleavage.

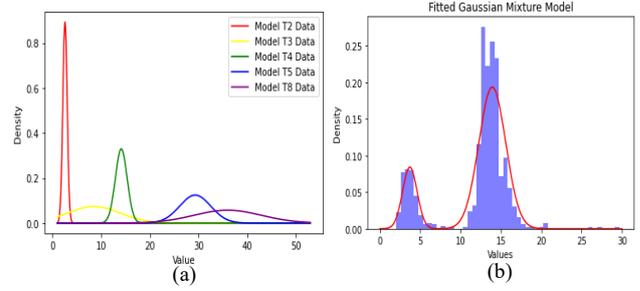
(a)        (b)
Fig. 7(a) 5-mode distribution (b) distribution of *t3*.

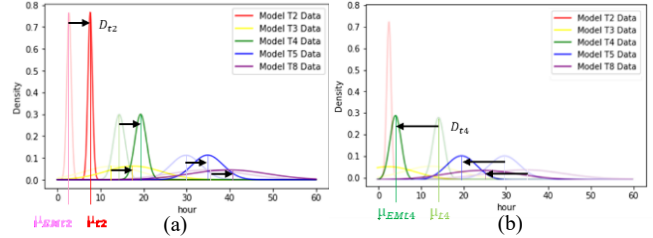
(a)        (b)
Fig. 8 Distributions after calibration (a) $D_{ti} > 0$ (b) $D_{ti} < 0$.

cleavage blastomeres with (**ID**: A0, A1, B0, B10, B11) and (*SizeS_4, ContS_4*)=(86%, 93%) and (SizeS_4, ContS_4)= (97%, 93%).

Fig. 6 shows the result of the irregular cleavage, where one blastomere is directly cleavaged into four blastomeres. Such irregular cleavaged blastomere is highlighted by the label "IRC" at the top right corner of the TLI images, as shown in Fig. 6(b). Fig. 7(a) shows the result of the local EM algorithm, i.e., the normal distribution of ACT in five categories: *t2, t3, t4, t5,* and *t8*. Where the mean values of *t2*, t3, t4, t5, t8 are 2.56, 8.49, 14.13, 29.34, 36.02 hrs, respectively, and the standard deviation of *t2, t3, t4, t5, t8* is 0.45, 5.38, 1.21, 3.21, 6.93 hrs, respectively. Fig. 7(b) shows that the initial cleavage time distribution for *t3* is modeled as GMM (Gaussian Mixture Model), the estimated result of t3 must be adjusted according to t4 with the help of the pre-trained EM (see the lower red arrow in Fig.1). Fig.8(a) and Fig.8(b) respectively shows the calibrated distributions for two different cases after applying Eq.(1) and Eq.(2).

*Classifier_2* can quickly annotate data and accurately distinguish between tPNa and tPNf. However, it is limited by data imbalance issues and performs poorly on highly complex images, often misclassifying *t6* and *t7* as *t8*. In contrast, training the *Segmentor* in Fig. 1 with single-class annotated data allows to effectively distinguish between *t6, t7*, and *t8*. Still, the data annotation process is time-consuming and cannot differentiate between tPNa and tPNf. Our method is characterized by utilizing the complementary effect of *Classifier_2* and *Segmentor*, combining them can achieve better results than just using either one of them. Additionally, incorporating EM can significantly enhance the prediction accuracy of CTP. Here, we compared the time points of cleavage stage using three sources: the manually annotated Ground Truth (GT) from Vitrolife™, and results from CTP, which was developed in this work. The average difference between GT and Vitrolife™ is 6 hrs, while CTP can reduce the average difference to less than 3 hrs.

Table 3 compares the predicted timepoints of three embryos using different methods. Embryo C experienced a direct cleavage, where the number of blastomeres directly increased from 1 to 4. In contrast, Vitrolife™ incorrectly identified non-existent *t2* and *t3* for Embryo C which actually underwent direct cleavage from one cell to four cells. In contrast, CTP correctly identified the t4 time point. Additionally, while Vitrolife™ provides only six cleavage time points, CTP further provides *t6* and *t7*. The CTP scheme designed with cost-effective EM-calibrated federated learning can give the complete timepoints of cleavage. It provides more accurate time points than other methods or commercial TLI device, which is crucial for analyzing blastocyst quality, implantation, or pregnancy rates.

Table 3 Comparison of ACT prediction of three embryos.

| TLI Image index | | tPNf | t2 | t3 | t4 | t5 | t6 | t7 | t8 |
|---|---|---|---|---|---|---|---|---|---|
| 1 | GT | 21.1 | 23.1 | 33.4 | 34.1 | 45.1 | 46.1 | 46.7 | 49.7 |
| | Vitrolife™ | 21.1 | 23.1 | 33.4 | 34.1 | 45.1 | N/A | N/A | 49.7 |
| | CTP (without EM) | 21.1 | 23.1 | N/A | 34.1 | 45.7 | N/A | N/A | 46.7 |
| | CTP | 21.1 | 23.1 | 33.4 | 34.1 | 45.1 | 46.1 | 47.4 | 50.1 |
| | Vitrolife™ Error | 0.00% | 0.00% | 0.00% | 0.00% | 0.00% | N/A | N/A | 0.00% |
| | CTP Error | 0.00% | 0.00% | 0.00% | 0.00% | 0.00% | 0.00% | 1.50% | 0.80% |
| 2 | GT | 25.2 | 27.7 | 38.4 | 39.1 | 51.1 | 51.6 | 51.8 | 66 |
| | Vitrolife™ | 25.4 | 27.9 | 38.4 | 38.9 | 50.5 | N/A | N/A | 51.8 |
| | CTP (without EM) | 25.2 | 27.7 | N/A | 39.2 | 52.3 | N/A | N/A | N/A |
| | CTP | 25.2 | 27.7 | 38.4 | 39.1 | 51.3 | 51.6 | 52 | 58.5 |
| | Vitrolife™ Error | 0.79% | 0.72% | 0.00% | 0.51% | 1.17% | N/A | N/A | 21.52% |
| | CTP Error | 0.00% | 0.00% | 0.00% | 0.00% | 0.39% | 0.00% | 0.39% | 11.36% |
| 3 | GT | 27 | pass | pass | 32 | 44.7 | N/A | N/A | N/A |
| | Vitrolife™ | 27 | 32 | 32.3 | 35 | 44.7 | N/A | N/A | N/A |
| | CTP (without EM) | 27 | 32 | pass | 32.6 | 44.7 | N/A | N/A | 65.6 |
| | CTP | 27 | pass | pass | 32 | 45.3 | 61.4 | N/A | 65.6 |
| | Vitrolife™ Error | 0.00% | N/A | N/A | 9.38% | 0.00% | N/A | N/A | N/A |
| | CTP Error | 0.00% | N/A | N/A | 6.86% | 1.34% | N/A | N/A | N/A |

## Conclusion

The major contributions of this study are threefold. First, with the simplicity of EM algorithm, the proposed federated learning framework can be easily adapted to data of small local clinics or large hospitals, lowering the barrier to assessing embryo quality and making it more feasible for various institutions to engage in research of reproductive medicine. Second, the framework can assist in obtaining more objective embryo parameters, identifying IRC embryos, and allowing embryologists to dedicate more time to studying exceptional cases. Finally, the proposed CTP scheme can be used as a practical tool to reduce the time and effort spent by embryologists manually observing embryo formation and acquiring important information regarding how the embryo develops. Another noteworthy point is that our framework can provide more accurate cleavage time points, in addition to the two time points t6 and t7 that are normally absent in the report of many TLI makers (e.g. Vitrolife™).

In the future, with more embryo images and annotated data collected from other medicine institutions, it is believed that performance of the proposed method can be further improved. By combining with morphokinetic parameters of other stages and patient's physiological data, the efficacy of ART can be enhanced, leading to higher embryo selection accuracy and increased implantation rates, thereby significantly increasing the success rates of infertility treatments. We note that the current framework cannot ideally deal with embryo images present with severe overlapping or complex backgrounds. Also, the performance of many deep learning models used in this study is inevitably influenced by the model size, hyperparameters setting, and regularization means. Finally, for some resource-limited areas or countries, affordable instrumental software is highly demanded. To this end, we will study the plausibility of incorporating the weigh-sharing design (Wang, et. al. 2022) into the network of *Classifier_1, Classifier_2*, *Segmentor*.